\documentclass[a4paper,10pt]{article}
	\usepackage{amsmath}
	\usepackage{graphicx}

\title{\Large Detailed analysis of an endoreversible fuel cell : Maximum
power and optimal operating temperature determination}
\author{A. Vaudrey\footnote{FEMTO-ST/ENISYS Institute, UMR CNRS
6174, University of Franche-Comte, Parc technologique, 2, avenue Jean
Moulin, 90 000 Belfort, France. Phone : +33 (0)3 84 57 82 27 / Fax : +33 (0)3 84
57 00 32. email : alexandre.vaudrey@femto-st.fr}, P.~Baucour, F.~Lanzetta, R.~Glises \\
\small{FEMTO-ST/ENISYS Institute, UMR CNRS 6174, University of Franche-Comte,
France.}}
\date{}

\begin{document}
\maketitle

\begin{abstract}
	Producing useful electrical work in consuming chemical energy, the fuel cell has
	to reject heat to its surrounding. However, as it occurs for any other type of
	engine, this thermal energy cannot be exchanged in an isothermal way in finite
	time through finite areas. As it was already done for various types of systems,
	we study the fuel cell within the finite time thermodynamics framework and
	define an endoreversible fuel cell. Considering different types of heat transfer
	laws, we obtain an optimal value of the operating temperature, corresponding to
	a potentially maximum produced electrical power. Finally, two fundamentals
	results are obtained : high temperature fuel cells could extract more useful
	power from the same quantity of fuel than low temperature ones with lower
	efficiencies. Thermal radiative exchanges between the fuel cell and its
	surrounding have to be avoided so far as possible, because of their negative
	effects on optimal operating temperature value.
\end{abstract}


\section{Introduction}

The fuel cell (FC) is usually described as a system that directly converts into
electricity the chemical energy provided by a process considered as a
combustion~\cite{Larminie}. Since it extracts a fraction of useful work from the
energy provided by a fuel and a combustive, it could then be also viewed as a
particular type of engine. Thereby, as any other engines and according to the
second law of thermodynamics, all the provided chemical energy could not be
converted into a useful form and a heat quantity is always rejected by the FC to
its surrounding. The minimum thermal energy released, corresponding to the
maximum produced work, is rejected by a reversible fuel cell (RFC) with no
internal production of entropy~\cite{RE-029-0179-0195,FDFC-2008}. In order to
include the FC into the finite time thermodynamics framework, we demonstrate in
the present study the equivalence between RFC and the Carnot heat engine (CHE),
system that extracts the maximum useful work from a given heat quantity.

As it was highlighted by Chambadal~\cite{Chambadal-tr-eng} and Novikov~\cite{Novikov}
and later by Curzon and Ahlborn~\cite{AJPh-043-0022-0024}, a reversible heat
engine can only operate in exchanging heat as an infinitely slow process and
therefore can not produce any practical useful power. In fact, energy or mass
transfers during finite durations or across finite areas are sources of entropy
and leads to a decrease of whole system performances. Taking into account these
irreversibilities in the analysis and optimization of system performances is the
scope of the finite time thermodynamics\cite{PECS-030-0175-0217} (FTT), also
called finite dimensions optimal thermodynamics\cite{entropy-011-0529-0547}
(FDOT). A system producing entropy only because of irreversible exchange
processes with its surrounding is usually qualified as endoreversible
\cite{PhRA-019-1272-1276,PhRA-019-1277-1289}. The concept of endoreversibility
has been successfully applied to a large scale of systems, including different
types of engines \cite{AJPh-053-0570-0573,AJPh-075-0169-0175}, heat
pumps~\cite{JAPh-102-034905}, chemical systems~\cite{JCPh-072-5118-5124},
distillation devices~\cite{JPhC-088-0723-0728}, or pneumatic
actuators~\cite{IJFP-003-0005-0012}.

As any other type of energy converters, the FC has to reject its own generated
heat flux across finite areas or during finite times and can not produce any
useful power when being entirely reversible. Indeed, to provide electrical work,
a FC, even reversible, has to release the heat produced by chemical process. To
be possible, this exchange must be based on a finite temperature difference that
produces entropy. Hence, considering a finite thermal conductance between a FC
and its surrounding, the Gibbs energy variation, usually sufficient to assess
its performances, is then not the only parameter influencing efficiency and
quantity of useful work produced. Taking into account the same limitations as
for other systems, the FC could be considered as an endoreversible system,
producing entropy only in rejecting heat flux in the
ambiance~\cite{FDFC-2008}. Relevant design and optimization methods of
thermodynamical machines, derived from FDOT, could then be applied to theses
particular systems~\cite{entropy-011-0529-0547}.

In this article, we study and carry out the performances (produced work and
energy efficiency) of an endoreversible fuel cell (EFC) considering different
types of heat exchange laws (linear or not). We highlight the existence of an
optimal operating temperature and discuss results dealing with different types
of hydrogen fuelled FC. Even if the present work focuses on EFC, a similar
analysis could be applied to a real, i.e. irreversible FC housing internal
productions of entropy.

\section{The endoreversible fuel cell}

\subsection{Reversible fuel cell and Carnot heat engine}

Let us consider an open and steady state system operating at constant
temperature $T$ and pressure $p$ and housing the followed exothermic chemical
reaction~: \begin{equation} 
    \sum_{j \in {\cal R}} \nu_{j} \cdot A_{j} \to \sum_{k \in {\cal
    P}} \nu_{k} \cdot A_{k} \label{chemreact}
\end{equation} with $A_{i}$ the chemical species and $\nu_{i}$ their
corresponding stoichiometric coefficients. ${\cal R}$ and ${\cal P}$ are
respectively the groups of reactants and products of reaction both considered as
ideal gases. Besides molar quantities of reactants and products, it is assumed
that the whole system exchanges work $W$ and heat $Q$ with its surrounding.

Considering a system (drawn on Fig.~1) noted 1 like e.g. a fuel cell, which
directly converts into work the chemical energy provided by reaction
(\ref{chemreact}), the energy and entropy balances are respectively :
\begin{align}
	\delta W_1 & = -\Delta h(T) \cdot {\rm d}\xi - \delta Q_1
	\label{first-law-01} \\
	\delta_{\rm i}S_1 & = \Delta s({\bf p},T) \cdot {\rm d}\xi +
	\frac{\delta Q_1}{T} \label{second-law-01}
\end{align} with $\Delta h$ and $\Delta s$ respectively the variations of
enthalpy and entropy through reaction (\ref{chemreact}), $\delta Q_1$ the
exchanged heat quantity, $\delta_{\rm i}S_1$ the internal production of entropy,
$\bf p$ the vector of the partial pressures of both reactants and products and
$\xi$ the reaction progress coordinate defined by : \begin{equation} 
	{\rm d} \xi = \left( - \frac{{\rm d} N_{j}}{\nu_{j}} \right)_{j \in
	{\cal R}} = \left( \frac{{\rm d} N_{k}}{\nu_{k}} \right)_{k \in {\cal
	P}}
\end{equation} Combining (\ref{first-law-01}) and (\ref{second-law-01}), the
variation of produced work can be expressed as : \begin{align} 
	\delta W_1 & = -\Delta h(T) \cdot {\rm d} \xi + T \cdot (\Delta s({\bf
	p},T) \cdot {\rm d}\xi -\delta_{\rm i}S_1) \nonumber \\
        & = -\Delta g({\bf p},T) \cdot {\rm d} \xi - T \cdot \delta_{\rm i}S_1
\end{align} with $\Delta g$ the variation of the Gibbs energy $g$ for chemical
process (\ref{chemreact}). \begin{figure}[h]
	\begin{center}
	\includegraphics{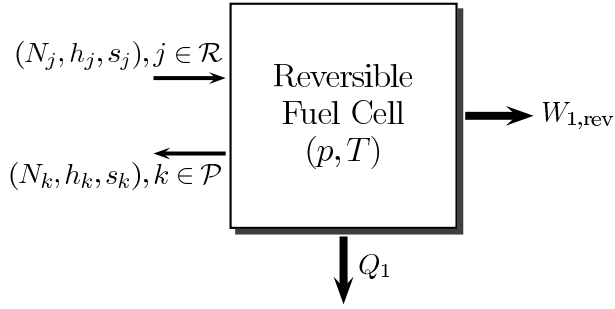}
	\caption{Schematic diagram of system $1$}
	\end{center} 
\end{figure}

Logically, the maximum value of work $\delta W_1$ will be produced by a
reversible system, i.e. with no internal production of entropy ($\delta_{\rm
i}S_1 = 0$) : \begin{equation} 
	\delta W_{1,{\rm rev}} = -\Delta g({\bf p},T) \cdot {\rm d} \xi
\end{equation} Dividing this provided work by the reaction progress, we can
express the molar reversible work $w_{\rm rev}$ as : \begin{equation} 
    w_{1,{\rm rev}} = \frac{\delta W_{1,{\rm rev}}}{{\rm d}\xi} = -\Delta g({\bf
    p},T) \label{work-01}
\end{equation} that corresponds to the quantity of useful energy provided by a
reversible system for one mole of consumed chemical reactants. The energy
efficiency $\eta$ will be defined as the fraction of useful energy $w_1$ on
consumed energy, e.g. for the considered reversible system~\cite{Larminie}~: \begin{equation} 
	\eta_{1,{\rm rev}} = \frac{w_1}{-\Delta h(T)} = \frac{\Delta g({\bf
	p},T)}{\Delta h(T)} \label{eff01}
\end{equation} Regarding to a real FC, the RFC would be characterised by the
absence of overpotentials which are direct results of internal production of
entropy~\cite{IJHE-033-4161-4170}. The well known polarization curve i.e. the voltage
vs. current produced by the FC should then be replaced by a constant value of
voltage, equal to the equilibrium one.

Let us consider now an other system, noted $2$ and drawn on Fig.~2, that
produces only heat from chemical reaction (\ref{chemreact}), which is used as a
hot source of a Carnot heat engine (CHE). \begin{figure}[h]
	\begin{center}
		\includegraphics{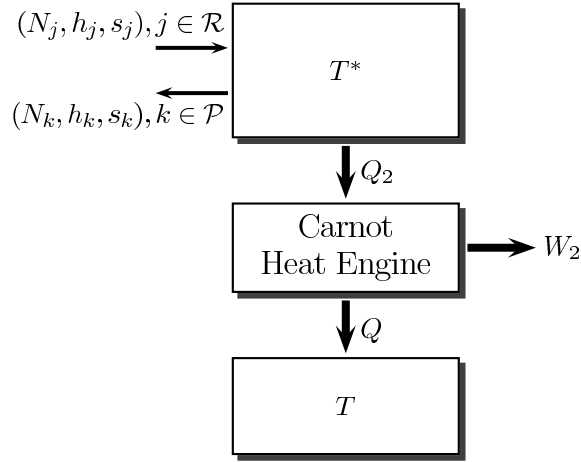}
		\caption{Schematic diagram of system $2$ feeding a CHE.}
	\end{center} 
\end{figure} First and second laws applied to this combustion system gives~:
\begin{align}
	0 & = -\Delta h(T) \cdot {\rm d}\xi - \delta Q_2 \label{first-law-02} \\
	\delta_{\rm i}S_2 & = \Delta s({\bf p},T) \cdot {\rm d}\xi +
	\frac{\delta Q_2}{T^{*}} \label{second-law-02}
\end{align} with $T^*$ the temperature of the chemical process. Supposing it
reversible ($\delta_{\rm i}S_2 = 0$) and combining (\ref{first-law-02}) and
(\ref{second-law-02}), we obtain the temperature $T^*$, sometimes called
\emph{entropic temperature} \cite{IJTS-040-0659-0668-ang} or \emph{equilibrium
temperature} \cite{These-deGroot-2004-eng}~: \begin{equation} 
	T^*({\bf p},T) = \frac{\Delta h(T)}{\Delta s({\bf p},T)}
	\label{entropicT}
\end{equation} This temperature corresponds to the one reached by a combustion
process as (\ref{chemreact}) and both consuming and rejecting reactants and
products at temperature $T$. Supplying a Carnot engine operating between
temperatures $T$ and $T^*$ with heat $\delta Q_2$ produced by previous ideal
combustion system, we can produce a work quantity
\cite{RE-029-0179-0195,FDFC-2008}~: \begin{align} 
	\delta W_{2,{\rm rev}} & = \delta Q_2 \cdot \eta_2 = \delta Q_2 \cdot
	\left(1 - \frac{T}{T^*({\bf p},T)} \right) = \underbrace{-\Delta
	h(T)}_{=q^*} \cdot {\rm d}\xi \cdot \left(1 - \frac{T \cdot \Delta
	s({\bf p},T)}{\Delta h (T)} \right) \nonumber \\
        & = -\Delta g({\bf p},T) \cdot {\rm d}\xi \label{work-02}
\end{align} It can be seen that $w_{1,{\rm rev}} = w_{2,{\rm rev}}$ and
$\eta_{1,{\rm rev}} = \eta_{2,{\rm rev}}$ and finally concluded on the
thermodynamical equivalence of both reversible systems \cite{FDFC-2008}, i.e. of RFC and CHE. Indeed, system $2$ is defined as a burner (housing an
exothermic chemical reaction) associated to a Carnot engine that producing work.
Efficiency $\eta_2$ of the whole system is the one of a classical ideal heat
engine.

Hence, a RFC could be viewed as a CHE operating with a heat source at
temperature $T^*$ and a cold sink at its own temperature $T$ (see Fig.~3a).

\subsection{Endoreversible fuel cell}

As originally demonstrated by Chambadal \cite{Chambadal-tr-eng} and
Novikov~\cite{Novikov}, an entirely
reversible engine can exhange thermal energies only in a infinitely slowly
process and finally can not produce any useful power. Practically, heat transfer
can not occur across finite exchange areas in an isothermal way and a finite
difference of temperature is necessary to allow rejection of produced heat. In
presented our model on Fig.~3b, the FC operates at temperature $T$ in an
ambiance at the cold temperature noted $T_c$. \begin{figure}[h]
	\begin{center}
		\includegraphics{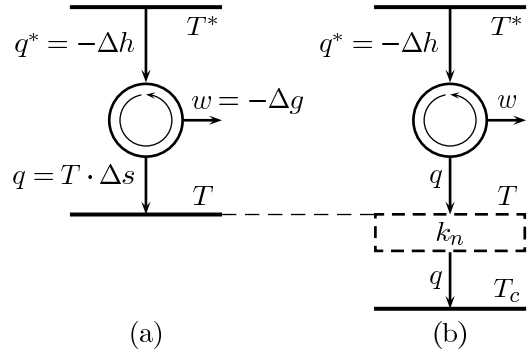}
		\caption{Equivalent heat engines of (a) reversible fuel cell
		(RFC) and (b) endoreversible fuel cells (EFC).}
	\end{center} 
\end{figure} Rejected molar heat quantity, noted $q$ is now a function of
temperatures $T$ and $T_c$. De Vos~\cite{AJPh-075-0169-0175} showed that different heat transfer
laws could be applied to this system. We will define the general heat transfer
law as~: \begin{equation} 
	q(T,T_c) = k_n \cdot (T^n - T_c^n) \label{conductance-01}
\end{equation} where $k$ is equivalent to a thermal conductance (usually defined
as a product of heat transfer coefficient and heat transfer area), but divided
by reaction progress as precedently did for molar work of relation
(\ref{work-01}). $n$ is an integer representative of the type of heat transfer
law. Then, the heat exchange process will be called linear if $n=1$ and
nonlinear if $n > 1$.

The equivalent heat engine of Fig.~3b being internally reversible (i.e.
endoreversible), its entropy balance is : \begin{equation} 
	\frac{q^*(T)}{T^*({\bf p},T)} = \frac{q(T,T_c)}{T}
	\label{balance-entropy-01}
\end{equation} with $q^{*}(T) = -\Delta h(T)$ the heat delivered by an
equivalent hot heat source (Fig.~2). Combining (\ref{work-02}),
(\ref{conductance-01}) and (\ref{balance-entropy-01}), the molar work becomes :
\begin{equation} 
	w  = q(T,T_c) \cdot \left( \frac{T^*({\bf p},T)}{T} - 1 \right) = k_n
	\cdot (T^n - T_c^n) \cdot \left( \frac{T^*({\bf p},T)}{T} - 1 \right)
	\label{work-03}
\end{equation} and related energy efficiency : \begin{equation} 
	\eta = 1 - \frac{T}{T^*({\bf p},T)} =\eta_{1,{\rm rev}} \label{eff-03}
\end{equation} Considering expression (\ref{work-03}) of the molar work $w$,
three cases have to be considered : \begin{enumerate}
	\item If $T = T_c$, according to equation (\ref{conductance-01}), no
		heat quantity can be exchanged and then $q = T \cdot \Delta s=0$
		(see Fig.~3a). Therefore, molar variation of entropy
		$\Delta s$ is null and chemical process can not occur. Thereby,
		the whole system can not produce any useful work and $w=0$.
	\item If $T$ increases, it can reach its maximum value, noted
		$T_{\max}$, that corresponds to an equality between the
		operating temperature and the entropic one, i.e. root of the
		following function : \begin{equation} 
			f(T) = T - T^*({\bf p},T)
    \end{equation} This situation leads to the vanishing of Gibbs energy
    variation : \begin{equation} 
	    -\Delta g({\bf p},T_{\max}) = 0
    \end{equation} and to $w=0$, according to (\ref{work-01}).
    \item If $T_c < T < T_{\max}$, the work $w$ is a continuous and positive
	    function of $T$ which could be maximized.
\end{enumerate} Unlike other endoreversible engines
\cite{entropy-011-0529-0547}, that are usually supposed to operate with
independent values of hot and cold temperatures, relationship (\ref{work-03})
(i.e. the molar produced work) depicts a non linear link between entropic and
operating temperatures, $T^{*}$ and $T$. This situation corresponds to an
equivalent heat engine whose value of the hot source temperature is a function
of the cold sink one. Consequently, the maximum power corresponding temperature
could not be given as the usual geometric mean of high and low temperatures
\cite{JAPh-079-1191-1218} $\sqrt{T_c \cdot T_{\max}}$. However, the optimal
temperature $\widehat{T}$ still corresponds to a maximum value of $w$ :
\begin{equation} 
	\widehat{T} = \arg \max_T \left( w(T) \right)
\end{equation} This relationship will be optimized by a Newton numerical
algorithm using $\sqrt{T_c \cdot T_{\max}}$ as an initial point which turns out
to be a good approximation (see Fig.~6).

\section{Hydrogen fuel cell}

As a practical example, let's consider the case of an endoreversible fuel cell
(EFC) operating in consuming hydrogen as fuel. Chemical reaction
(\ref{chemreact}) is : \begin{equation} 
	{\sf H}_{2} + 1/2 \cdot {\sf O}_{2} \to {\sf H}_{2}{\sf O}
	\label{chemreact-H2}
\end{equation} The surrounding temperature is fixed for example at $T_c =
298,15\,{\rm K}$. Thanks to experimental correlations published by the
NIST~\cite{NIST-Thermo}, we have represented in Fig.~4 the evolution of the entropic
temperature $T^{*}$ vs. reduced temperature for pure oxygen as combustive :
\begin{equation}
	\theta = (T-T_c)/(T_{\max}-T_c) \label{reduced-temp}
\end{equation} Corresponding maximum temperature for pure ${\sf O}_2$ is
$T_{\max}(x_{{\sf O}_2} = 1) \simeq 4\,310\,{\rm K}$. \begin{figure}[h!]
	\begin{center}
		\includegraphics[width=9cm]{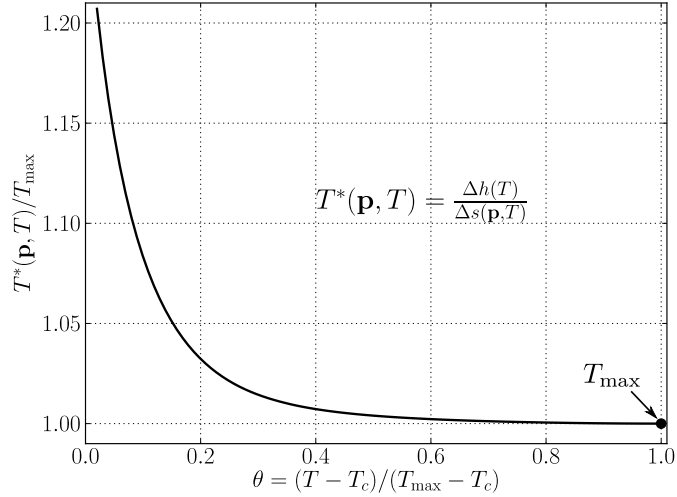}
	\end{center}  
	\caption{Entropic temperature $T^*$ of relation
	(\ref{entropicT}) regarding to the fuel cell (FC) operating one
	$T$ for pure ${\sf O}_2$ as combustive ($x_{{\sf O}_2}=1$).}
\end{figure} Using air as combustive leads to a similar curve, but with $T_{\max}(x_{{\sf O}_2}=0,21)
\simeq 3\,884\,{\rm K}$. Dot at the end of curve represents the maximum
temperature, corresponding to $\theta = 1 \Rightarrow T^*=T_{\max}$. Using
relation (\ref{eff01}) and (\ref{eff-03}), energy efficiencies for some pure
oxygen and air as combustive are represented on Fig.~5, with values of
coordinate $\theta$ based on maximum temperature with pure oxygen
$T_{\max}(x_{{\sf O}_2} = 1)$.

It can be observed on Fig.~5 a quasi linear decrease of $\eta$ with the
operating temperature $T$, leading to a zero value for $T = T_{\max}$ with pure
oxygen as combustive. Energy efficiency corresponding to air consumption reach
zero at is own value of $T_{\max}$, less than the one of pure oxygen. Range of
$\theta$ values does not start at zero because of the possible liquid water
formation that can occur below $100^{\circ}{\rm C}$. \begin{figure}[t!]
	\begin{center}
		\includegraphics[width=9cm]{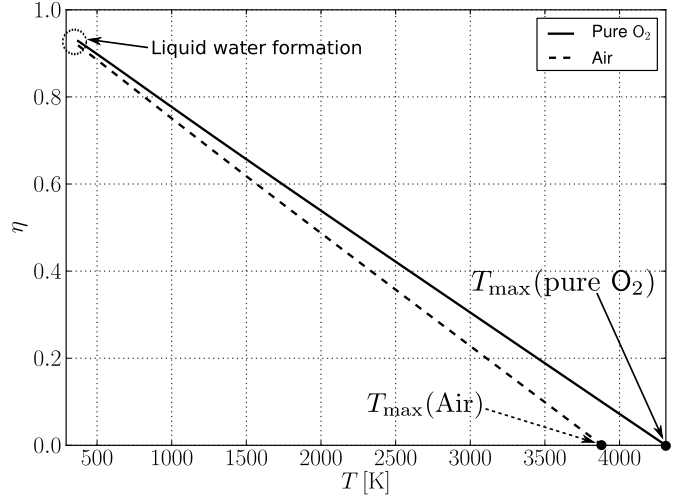}
		\caption{Energy efficiencies $\eta$ of RFC (relation
		(\ref{eff01}) and (\ref{eff-03})) fed by pure ${\sf O}_2$ (full
		line) and air (dashed line) as combustive. Minimum values of
		$\eta$ for both cases corresponds to respective values of
		$T_{\max}$.}
	\end{center} 
\end{figure}

\subsection{Linear heat transfer law}

Considering the molar work (\ref{work-03}) produced with linear transfer law
($n=1$), it can be given by : \begin{equation}
	w = k_1 \cdot (T - T_c) \cdot \left( \frac{T^*({\bf p},T)}{T}-1\right) =
	k_1 \cdot (T^*({\bf p},T) - T) \cdot \left(1 - \frac{T_c}{T} \right)
	\label{work-04}
\end{equation} That corresponds to the work produced by a Carnot heat engine
(CHE) operating between temperatures $T_c$ and $T$ and consuming some molar heat
$q = k_1 \cdot (T^*({\bf p},T) - T)$. Numerical results are drawn on Fig.~6.
\begin{figure}[t!]
	\begin{center}
		\includegraphics[width=9cm]{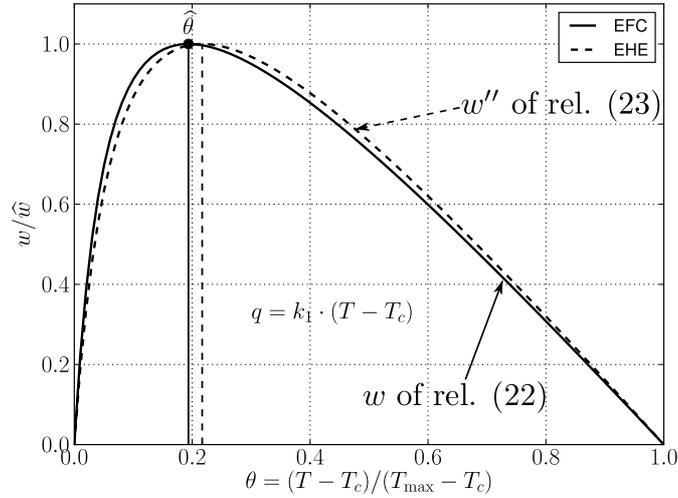}
		\caption{Reduced molar work $w/\widehat{w}$ vs. $\theta$ with
		the finite conductance (\ref{conductance-01}) for RFC and for an
		endoreversible heat engine (EHE) operating between constants
		temperatures $T_c$ and $T_{\max}$.}	
	\end{center} 
\end{figure} We have presented in continuous line (curve $1$) the evolution of reduced work
$w/\widehat{w}$ vs. reduced operating temperature $\theta$ of relation
(\ref{reduced-temp}). $\widehat{w}$ is the maximum value of $w$ regarding to
temperature $T$. The second dashed line (curve $2$), corresponds to the case
where hot temperature is supposed constant and equal to the highest one
$T_{\max}$ and where expression of the molar work $w''$ corresponds to those of
an usual endoreversible heat engine (EHE)~:\begin{equation} 
	w'' = k_1 \cdot (T_{\max} - T) \cdot \left( 1-\frac{T_c}{T}\right)
\end{equation} This engine is represented on Fig.~3b by replacing variable
high temperature $T^*$ by the constant one $T_{\max}$. In this particular case, it is easy to show that :
\begin{align} 
	\frac{\partial w''}{\partial T} & = k_1 \cdot \left( \frac{T_c \cdot
	T_{\max}}{T^2}-1 \right) = 0 \nonumber \\
	& \Rightarrow \widehat{T} = \sqrt{T_c \cdot T_{\max}} \label{geom-mean}
\end{align} and the optimal operating temperature $\widehat{T}$ also leads to a
maximum work : \begin{equation} 
	\widehat{w} = k_1 \cdot \left( \sqrt{T_{\max}}-\sqrt{T_c} \right)^2
\end{equation} and a related energy efficiency : \begin{equation} 
	\widehat{\eta} = 1 - \frac{\widehat{T}}{T_{\max}} = 1 -
	\sqrt{\frac{T_c}{T_{\max}}}
\end{equation} The difference between curves $1$ and $2$ is a direct consequence
of nonlinear link between the entropic temperature $T^*$ and the operating one
$T$. Therefore, the value of the optimal temperature $\widehat{T}$ corresponding
to $w = \widehat{w}$ is lightly modified and actually different from those
obtained by a geometric mean (\ref{geom-mean}) of both high and low
temperatures.  Main results are presented in Table~1.

\begin{table}
	\begin{center}
		\begin{tabular}{|c|cc|}
			\hline
			High temperature & $T^*({\bf p},T)$ & $T_{\max}$ \\
			\hline
			\Big.$\widehat{T}$ & $992\,{\rm K}$& $1\,076\,{\rm K}$ \\
			$\widehat{\eta}$ & $75,3\%$ & $72,3\%$\\
			\hline
		\end{tabular}
	\end{center}
	\caption{Maximum power operating temperature $\widehat{T}$ and related
	energy efficiency $\widehat{\eta}$ for variable and constant equivalent
	high temperatures.}
\end{table} The well used curve $w/\widehat{w} = f(\eta)$ is drawn on Fig.~7
for each endoreversible system. The difference between curves, due to the
nonlinearity of $T^*$ is here more significant, because of different values of
maximum efficiencies. The previous curves are useful for making the difference
between low and high-temperatures FC. We have carried out in Fig.~8 the
typical range of operating temperatures of protons exchange membrane fuel
cells, i.e. $60^{\circ}{\rm C} \leq T \leq 120^{\circ}{\rm C}$ \cite{Larminie}.
As previously presented on Fig.~5, a low-temperature fuel cell is characterized
with high value of its energy efficiency. \begin{figure}[t!]
	\begin{center}
		\includegraphics[width=9cm]{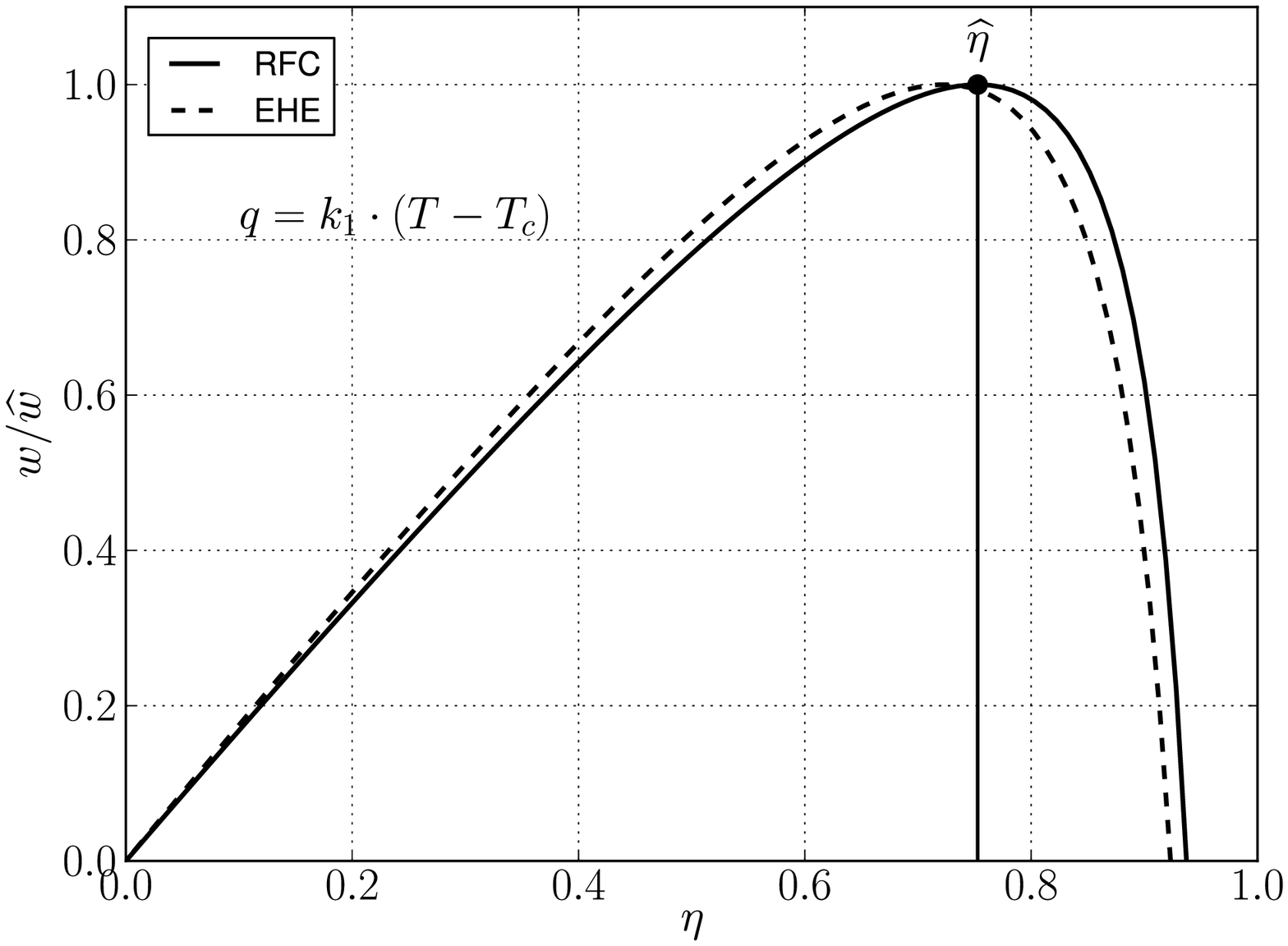}
	\end{center} 
	\caption{Reduced molar work $w/\widehat{w}$ regarding to energy
	efficiency $\eta$ for an EFC and for an EHE operating between $T_c$ and
	$T_{\max}$.}
	\begin{center}
		\includegraphics[width=9cm]{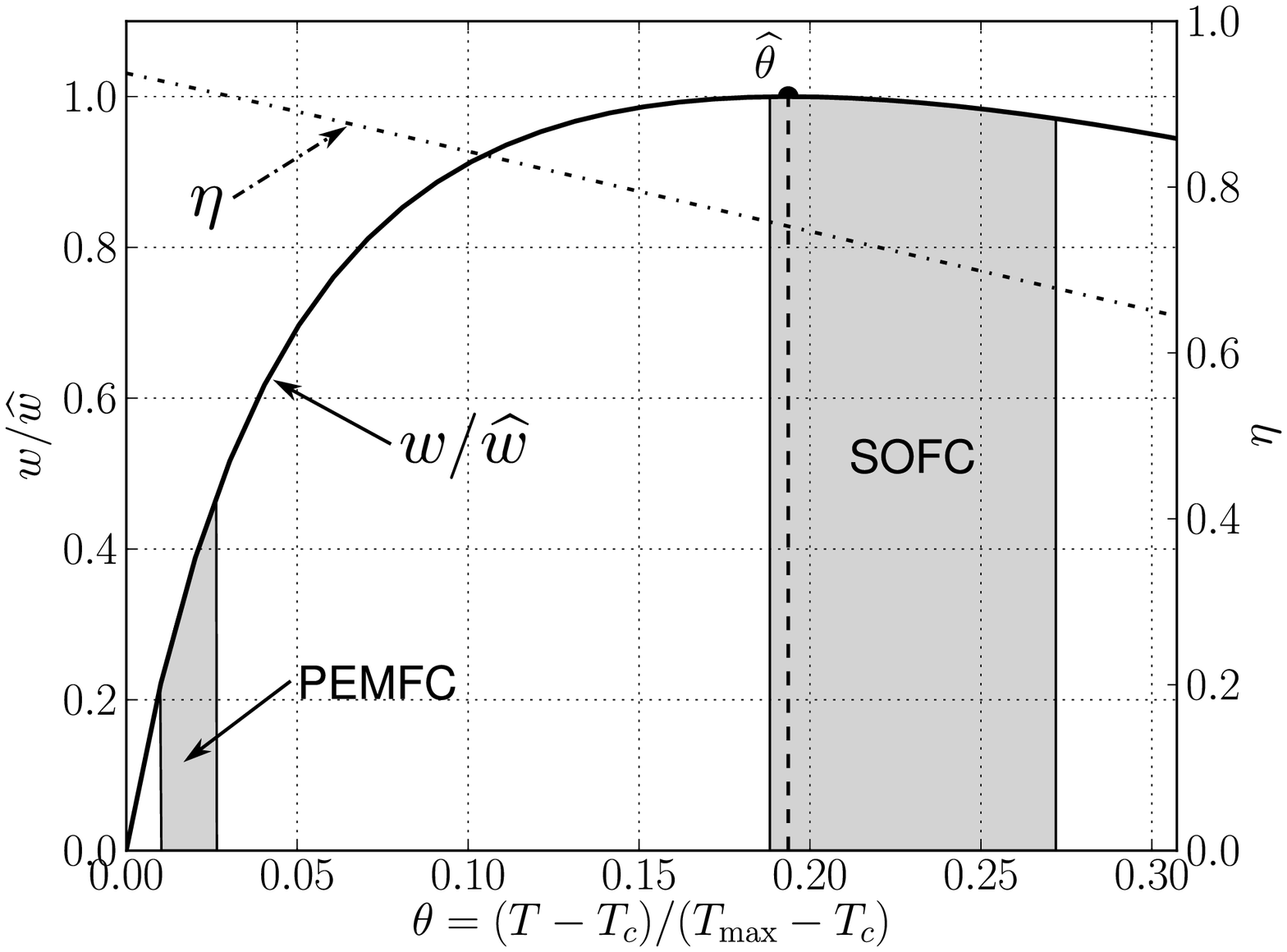}
	\end{center}
	\caption{Reduced molar work and efficiency of different types of FC for
	air as combustive.}
\end{figure} 
On a first hand, its low temperature difference with surrounding
prevents to reject important heat quantity $q$, and according to the Carnot
principle, to produce important work $w$ as presented on Fig.~8.
 On a second hand, high-temperature fuel cells, such as solid oxyde
fuel cells (SOFC), can easily evacuate generated thermal power, because of their
high temperature differences with the ambiance, and are also able to produce
high values of electrical power. On the right part of Fig.~8 is also drawn the
typical operating temperature range of SOFC, i.e. $700^{\circ}{\rm C} \leq T
\leq 1\,000^{\circ}{\rm C}$. However, this advantage is counterbalanced by a
lower energy efficiency, as shown on right part of Fig.~5.

This first result is interesting but does not take into account the difference
of heat transfer types between low and high temperature systems. Hence, high
temperature heat transfer processes are often driven by radiative effects.  

\subsection{Nonlinear heat transfer law}

Let's consider now the case of a radiative heat transfer phenomenon ($n=4$)
between the FC and the surrounding, we can express the produced molar work as :
\begin{equation} 
	w = k_4 \cdot (T^4 - T_c^4) \cdot \left( \frac{T^*({\bf p},T)}{T} - 1
	\right)
\end{equation} As presented on Fig.~9 (full line), the nonlinear heat exchange
law leads to a strong increase of the optimal temperature~: $\widehat{T} \simeq
2\,910\,{\rm K}$ in the same chemical conditions as previously.
\begin{figure}[h] 
	\begin{center}
		\includegraphics[width=9cm]{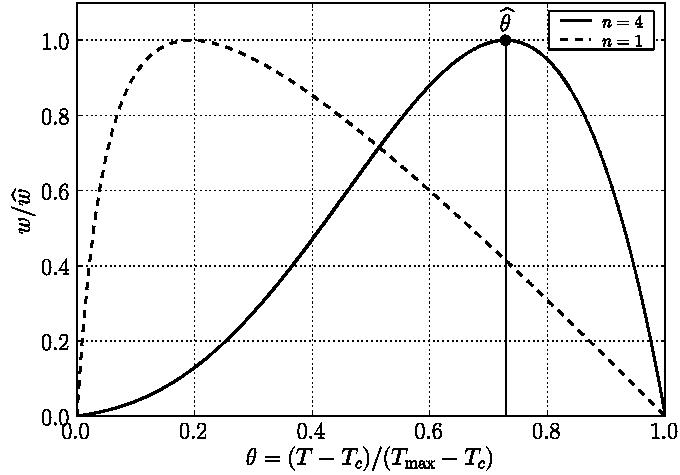}
	\end{center} 
	\caption{Evolutions of reduced molar work $w/\widehat{w}$ vs. $\theta$
	for $n=4$ (continuous curve) and $n=1$ (dashed curve).}
\end{figure} This result could be explained by the fact that the same heat
quantity needs a higher value of the temperature difference to be exchanged by a
radiative way than thanks to convective phenomenon. The molar heat $q$
corresponding to a maximum produced work could be only released beyond a
temperature difference limit higher with some radiative exchanges than with only
a convective phenomenon. To make the difference with previous linear case, we
have drawn on the same graph the molar work corresponding to the linear heat
transfer law (dashed line). The efficiency-work curves of both linear and
nonlinear heat transfers laws are presented on Fig.~10. \begin{figure}[h]
	\begin{center}
		\includegraphics[width=9cm]{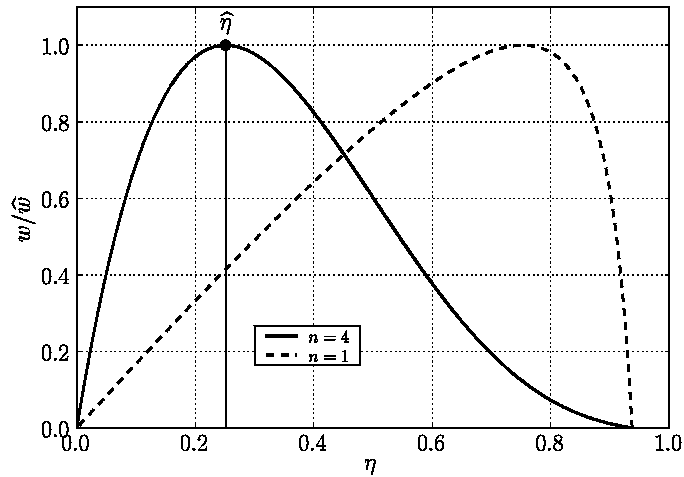}
	\end{center}
	\caption{Evolutions of reduced molar work $w/\widehat{w}$ regarding to
	energy efficiency $\eta$ with $n=4$ (full line) and $n=1$ (dashed
	line).}
\end{figure} Simultaneously to the increase of the maximum work operating
temperature, we can note a strong decrease of related energy efficiency :
$\widehat{\eta} \simeq 25\%$. And yet, as presented on Fig.~9, an increase of
the operating temperature corresponds to a decrease of its efficiency.  Hence,
the radiative heat transfer process appears to be unfavourable regarding to EFC
performances because it moves forward the optimal operating temperature from
those corresponding to convective heat exchange phenomena. 

Finally, a relevant thermal management system for high temperature FC will have
to minimize radiative losses in order  to decrease as much as possible the
optimal temperature value $\widehat{T}$ and to get the practical operating
temperature $T$ close to it.  

\section{Conclusions}

The formal equivalence between a reversible fuel cell (RFC) and a Carnot heat
engine (CHE), both supplied by the same reversible combustion processes, allowed
us to describe the former with the finite time thermodynamics approach.  The
main result is the definition of an endoreversible fuel cell (EFC), operating in
a reversible way but exchanging irreversibly heat with its surrounding, through
finite thermal conductances.
The optimization of the produced work regarding to the fuel cell operating
temperature has led to an optimal EFC temperature, numerically calculated for an
hydrogen-air reaction for standard conditions of pressure. The influence of the
heat transfer law type (linear or not) has been investigated for convective and
radiative cases. The last one seems to be unfavourable on the fuel cell
performances, because of the increase of the maximum produced work corresponding
temperature, compared to linear heat transfer case. At the same time, the
efficiency of such system decreasing with temperature, the maximum produced work
corresponding efficiency has also strongly decreased.  We can conclude on the
necessity for a relevant thermal management system to avoid radiative thermal
effects and to favour the convective heat exchange phenomena.

The present endoreversible fuel cell is based on an only one thermal finite
conductance due to the heat flux exchange with the ambiance. It would be
significant to also consider a non reversible chemical reaction, using for
example the results of chemical thermodynamics in finite
time \cite{ACR-017-0266-0271}. In the same way, different types of internal entropy
production could be progressively taken into account. 

Moreover, design and optimization processes of fuel cell systems have also to
take into account the fundamental Carnot principles. The heat flux rejected from
the system to the surrounding is fundamental and strongly influences at least
the electrical power produced.

\section*{Acknowledgements}

The authors would like to thank the French National Research Agency
(ANR), in the scope of its national action plan for hydrogen (PAN-H) for
financially supporting this work.

\section*{Nomenclature}

\begin{center}
\begin{tabular}{cl}
	& {\bf Acronyms} \\
	CHE & Carnot heat engine \\
	EFC & Endoreversible fuel cell \\
	EHE & Endoreversible heat engine \\
	FC & Fuel cell \\
	FDOT & Finite dimensions optimal thermodynamics \\
	FTT & Finite time thermodynamics \\
	RFC & Reversible fuel cell \\
	& {\bf Notations} \\
	$A$ & Chemical species \\
	$g$ & Molar Gibbs energy $[{\rm J/mol}]$ \\
	$h$ & Molar enthalpy $[{\rm J/mol}]$ \\
	$\nu$ & Stoichiometric coefficient \\
	$p$ & Pressure $[{\rm Pa}]$ \\ 
	$\cal P$ & Products of a chemical reaction \\
	$Q$ & Heat $[{\rm J}]$ \\
	$q$ & Molar heat quantity $[{\rm J/mol}]$ \\
	$\cal R$ & Reactants of a chemical reaction \\
	$s$ & Molar entropy $[{\rm J/(mol \cdot K)}]$ \\
	$\delta_{\rm i} S$ & Internal production of entropy $[{\rm J/K}]$ \\
	$T$ & Temperature $[{\rm K}]$ \\
	$W$ & Work $[{\rm J}]$ \\
	$w$ & Molar work $[{\rm J/mol}]$ \\
	$\xi$ & Reaction progress $[{\rm mol}]$ 
\end{tabular}
\end{center}

\bibliographystyle{unsrt}
\bibliography{/home/avaudrey/Sciences/References/Bibliographie}

\end{document}